\documentclass[10pt,conference]{IEEEtran}
\IEEEoverridecommandlockouts
\usepackage{cite}
\usepackage{textcomp}
\usepackage{xcolor}
\usepackage{stfloats}
\usepackage{amsmath,amssymb,amsfonts}
\usepackage{algorithmic}
\usepackage{graphicx}
\usepackage{textcomp}
\usepackage{xcolor}
\usepackage{multirow}
\usepackage{array}
\usepackage{colortbl}
\usepackage{verbatim}
\usepackage{url}
\usepackage{tcolorbox}
\usepackage[linesnumbered,lined,boxed,commentsnumbered]{algorithm2e}
\usepackage{booktabs}
\usepackage{listings}
 \usepackage{fancyhdr}
 \usepackage{verbatim}
 \usepackage{flushend}

\usepackage{pifont}
\newcommand{\tool}{\texttt{SOTitle}}

\newcommand{\DeepCom}{\texttt{Hybrid-DeepCom}}
\newcommand{\BART}{\texttt{BART}}
\newcommand{\CodeQue}{\texttt{Code2Que}}
\newcommand{\Transformer}{\texttt{Transformer}}
\newcommand{\NMT}{\texttt{NMT}}
\newcommand{\BM}{\texttt{BM25}}

\begin{document}

% \title{{\tool}: A Fine-grained Pseudo-code Generation Method via Transformer and Code Feature Extraction}

\title{{\tool}: A Transformer-based Post Title Generation Approach for Stack Overflow}

\author{\IEEEauthorblockN{Ke Liu\IEEEauthorrefmark{2},  
Guang Yang\IEEEauthorrefmark{2}, 
Xiang Chen\IEEEauthorrefmark{2}\IEEEauthorrefmark{3}\IEEEauthorrefmark{1},  Chi Yu\IEEEauthorrefmark{2}}
\IEEEauthorblockA{\IEEEauthorrefmark{2}\textit{School of Information Science and Technology},
\textit{Nantong University}, China\\
\IEEEauthorrefmark{3}\textit{State Key Laboratory of Information Security},\textit{ Institute of Information Engineering, Chinese Academy of Sciences}, China\\
Email: aurora.ke.liu@outlook.com, novelyg@outlook.com, xchencs@ntu.edu.cn, yc\_struggle@163.com}
}

% \author{Anonymous Author(s)}

% \author{\IEEEauthorblockN{Guang Yang}
% Nantong, China \\
% 1930320014@stmail.ntu.edu.cn}
% \and
% \IEEEauthorblockN{Xiang Chen}
% Nantong, China \\
% xchencs@ntu.edu.cn}
% \and
% \IEEEauthorblockN{Zhanqi Cui}
% Nanjing, China \\
% czq@bistu.edu.cn}
% \and
% \IEEEauthorblockN{Ke Liu}
% Nantong, China \\
% 806464561@qq.com}
% \and
% \IEEEauthorblockN{Chi Yu}
% Nantong, China \\
% yc\_struggle@163.com}
% \and
% \IEEEauthorblockN{Jinxin Cao}
% Nantong, China \\
% alfred7c@ntu.edu.cn}
% }

\maketitle

\begingroup
\renewcommand{\thefootnote}{}
\footnotetext[1]{\IEEEauthorrefmark{1} Xiang Chen is the corresponding author.}
\endgroup

\begin{abstract}

On Stack Overflow, developers can not only browse question posts to solve their programming problems but also gain expertise from the question posts to help improve their programming skills. Therefore,  improving the quality of  question posts in Stack Overflow  has attracted the wide attention of researchers. 
A concise and precise title can play an important role in helping developers understand the key information of the question post, which can improve the post quality.
However, the quality of the generated title is not high due to the lack of professional knowledge related to their questions or the poor presentation ability of developers.
A previous study aimed to automatically generate the title by analyzing the code snippets in the question post. However, this study ignored the useful information in the corresponding problem description.
Therefore, we propose an  approach {\tool}  for automatic post title generation by leveraging the code snippets and the problem description in the question post (i.e., the multi-modal input).
{\tool} follows the Transformer structure, which can effectively capture long-term dependencies through a multi-head attention mechanism. 
To verify the effectiveness of {\tool}, we construct a large-scale high-quality corpus from Stack Overflow, which includes 1,168,257 high-quality question posts for four popular programming languages.
Experimental results show that {\tool} can significantly outperform six state-of-the-art baselines in both automatic evaluation and human evaluation. To encourage follow-up studies, we make our corpus and approach publicly available.

\end{abstract}


\section{Introduction}
\label{sec:intro}

On Stack Overflow, developers can post their problems and wait for other community members to give corresponding answers to their problems. 
When developers encounter similar problems, such problems and corresponding answers are valuable and can be reused. 
Until now, millions of developers use Stack Overflow to search for high-quality answers for their programming problems. Moreover, Stack Overflow becomes a knowledge base for developers to learn programming skills by browsing high-quality posts~\cite{gao2020generating}\cite{cao2021automated}\cite{liu2021characterizing}\cite{chen2019sethesaurus}.

While the number of question posts in Stack Overflow has been growing rapidly, there are still a large number of problems, which have not received high-quality answers. These problems  may be unclear, unspecific, difficult to understand, or unattractive to related developers~\cite{asaduzzaman2013answering}.

%\cx{Analyze the previous studies on improve post quality...}

These low-quality question posts not only fail to get effective help, but also hinder the process of knowledge generation~\cite{anderson2012discovering}\cite{jin2019edits}.
To solve this problem, previous studies have been conducted on the quality prediction of the question posts~\cite{duijn2015quality,arora2015good ,gao2020generating,correa2013fit,trienes2019identifying,yao2013want,zhang2018code}. For example, Correa and Sureka~\cite{correa2013fit} surveyed the closed problems in Stack Overflow and found that high-quality problems should contain enough code for others to reproduce the problem.       
%\cx{focus on title quality problem, the root cause of title quality}

Except for the prediction of low-quality question posts and how to improve the performance of the quality prediction models, some studies~\cite{correa2013fit,toth2019towards,trienes2019identifying} found that an important reason for low-quality question posts is that developers do not create informative post titles.
Some developers may lack knowledge  related to their problems, or they may have poor presentation ability. For these developers, writing high-quality problem titles is a challenging task. 
Therefore, automatically generating post titles for Stack Overflow is needed.

%\cx{Introduce Gao et al. study}

\begin{figure}[htbp]
	\centering
    \vspace{-1mm}
	\includegraphics[width=0.5\textwidth]{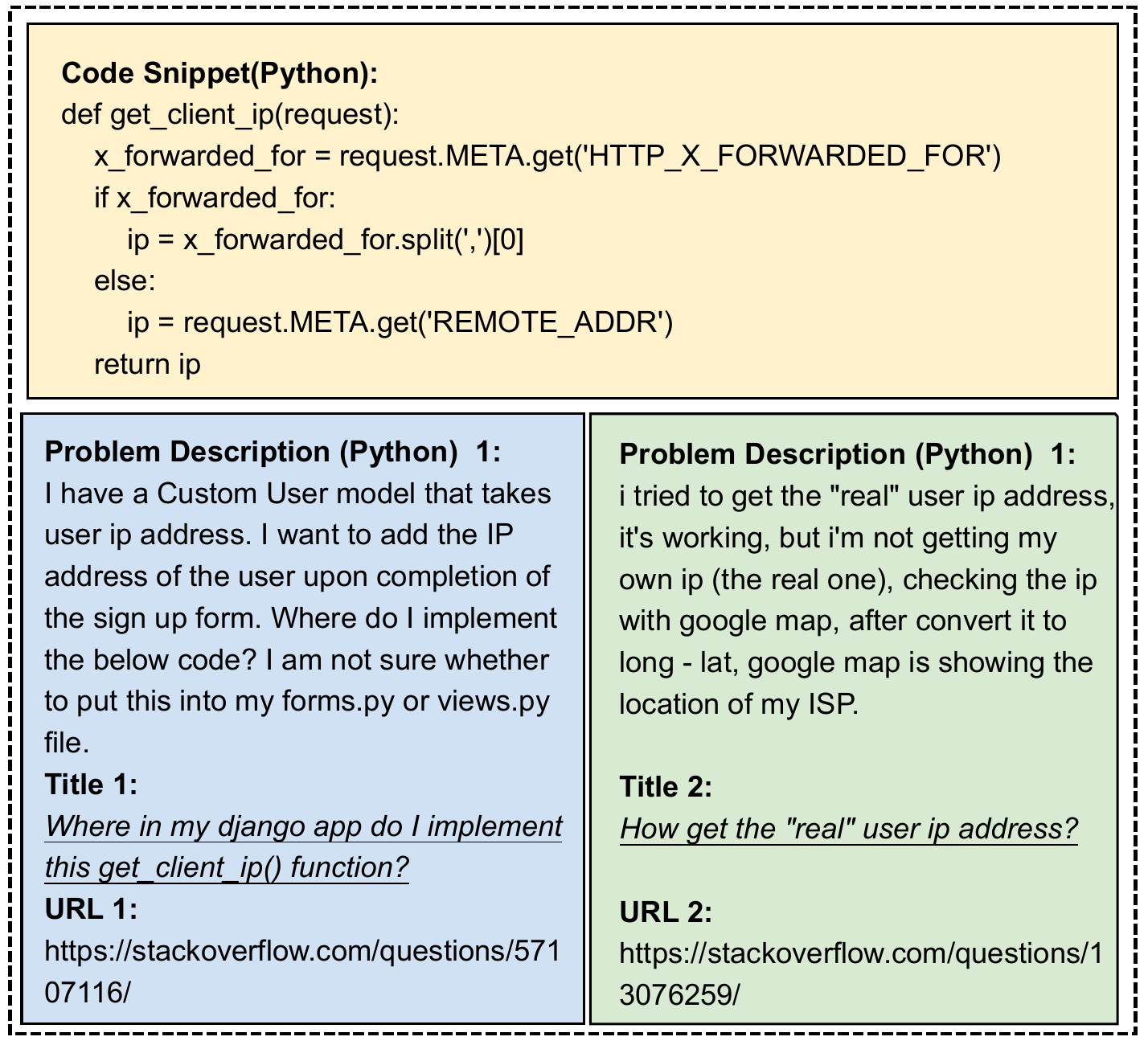}
	\caption{Two posts from Stack Overflow, which have the same code snippet but different problem descriptions}
    \vspace{-1mm}
	\label{fig:sameSnippet}
\end{figure}

To our best knowledge, Gao et al.~\cite{gao2020generating} were the first to automatically generate a post title according to a specific code snippet. However, they ignored the valuable information in the corresponding problem description. As shown in \figurename~\ref{fig:sameSnippet},
there are two posts in Stack Overflow, which have the same code snippet.
However, these two posts have different problem descriptions, which result in different post titles. 
According to this example, we can find the problem description of the post can provide valuable information, which sometimes cannot be provided by the code snippet for title generation. Therefore, considering both modalities (i.e., code snippet and problem description) as the input can help generate high-quality post titles.

The main ideas of our proposed approach {\tool} have three aspects. 
(1) We model the multi-modal input  by concatenating the code snippet and the problem description together and follows the Transformer structure.
% or using different encoders~\cite{hu2020deep} to handle two modalities respectively.
(2)  We formalize question post title generation for each programming language as separate but related tasks. Since related tasks can improve each other's performance by using shared and complementary information, we utilize multi-task learning~\cite{zhang2017survey}, which can guarantee the generalization of our approach by learning multiple tasks simultaneously.
(3) We use the method SentencePiece~\cite{kudo2018sentencepiece} to split the code snippet and the problem description to alleviate the OOV (out-of-vocabulary) issue in the post title generation problem. SentencePiece is designed for neural machine translation and it is a language-independent subword tokenizer
and detokenizer. 

Since there is no readily available corpus, to verify the effectiveness of our proposed approach, we gathered 1,168,257 high-quality problem posts for four different programming languages (i.e., Java, C\#, Python, and JavaScript) from Stack Overflow. 
Furthermore,  we compare {\tool} with six state-of-the-art baselines via automatic evaluation (i.e., Rouge~\cite{lin2004rouge}) and human study.
Specifically,
we first select the state-of-the-art post title generation approach {\CodeQue} proposed by Gao et al.~\cite{gao2020generating} as the first baseline. Then we select five approaches from source code summarization and text summarization domains (i.e., BM25~\cite{robertson2009probabilistic}, NMT~\cite{jiang2017automatically}, Hybrid-DeepCom~\cite{hu2020deep}, Transformer~\cite{vaswani2017attention}, and BART~\cite{lewis2019bart}) as the remaining baselines. 
Results of experimental study and human study show that {\tool} can generate higher-quality titles than these baselines.

To our best knowledge, the main contributions of our study can be summarized as follows.

\begin{itemize}
  \item We propose an approach {\tool} to generate the title of the question post from Stack Overflow by considering both code snippet and problem description. Specifically, we combine the code snippet and problem description as the multi-modal input of the Transformer structure. Then  we formalize  post title generation
for each programming language as separate but related tasks and utilize multi-task learning~\cite{zhang2017survey}. Finally, we use the SentencePiece method~\cite{kudo2018sentencepiece} to split the code snippet and the problem description to alleviate the OOV issue.
  
 \item We construct a high-quality corpus and this corpus contains 1,168,257 high-quality problem posts for four popular programming languages.
  
 \item  We conduct empirical studies on our construct corpus to compare {\tool} with six state-of-the-art baselines, including the recent  post title generation approach  {\CodeQue}~\cite{gao2020generating}. Final empirical results show the competitiveness of {\tool}. Moreover, we verify the effectiveness of {\tool} via human study.

  \item We develop a browser plugin based on our proposed approach. By using this plugin, users can generate high-quality titles for their submitted question posts in Stack Overflow. 

  \item We share our scripts, trained model, browser plugin, and corpora on our project homepage\footnote{\url{https://github.com/NTDXYG/SOTitle}}, which can facilitate the replication of our study and encourage more follow-up studies on the post title generation for Stack Overflow.
\end{itemize}

The rest of this paper is organized as follows.
Section~\ref{sec:related} analyzes related studies for post quality analysis, deep learning-based source code summarization, and deep learning-based text summarization. 
Section~\ref{sec:approach} introduces the framework of {\tool} and details of each component. 
Section~\ref{sec:setup} and Section~\ref{sec:results} show the experimental setup and results of empirical study and human study. 
Section~\ref{sec:threat} discusses the main threats to the effectiveness of our empirical study. 
Section~\ref{sec:conclusion} summarizes our study and shows several possible future directions.

\section{Related Work}
\label{sec:related}

In this section, we first summarize the related work for post quality analysis on Stack Overflow.
Since our approach aims to generate the post title by analyzing both the code snippet and the problem description, we also analyze the related work for deep learning-based source code summarization and text summarization.
Finally, we emphasize the novelty of our study.

\subsection{Post Quality Analysis on Stack Overflow}

%\cx{First Post quality is important.Then analyze the main problems of post quality analysis and related work.Finally, analyze the title quality problem and realted work.}

Improving post quality is an important research topic in Stack Overflow mining. 
% Previous work has examined the quality of problems in Stack Overflow~\cite{toth2019towards}\cite{correa2013fit}\cite{arora2015good}\cite{nasehi2012makes}\cite{yao2013want}\cite{ponzanelli2014understanding}\cite{trienes2019identifying}\cite{gao2020generating}. 
For example, 
Correa and Sureka~\cite{correa2013fit} investigated the closed questions in Stack Overflow and found that a good question should contain enough code for others to reproduce the problem.
Nasehi et al.~\cite{nasehi2012makes} performed qualitative assessment manually and investigated the important features of precise code examples in 163 Stack Overflow post answers.
Yao et al.~ \cite{yao2013want} found that the number of edits to a problem is a good indicator of the problem's quality.
Arora et al.~\cite{arora2015good} proposed a new approach to improve the accuracy of the question quality prediction model by using the content extracted from similar historical question posts.
Trienes et al.~\cite{trienes2019identifying} studied the approach of identifying unclear problems in Stack Overflow.
Ponzanelli et al.~\cite{ponzanelli2014understanding} developed an approach to automatically classify problems based on their quality.
Gao et al.~\cite{gao2020generating} were the first to use the sequence-to-sequence learning approach to automatically generate the post title based on the code snippet, which could improve the quality of the question post.

\subsection{Deep Learning-based  Text Summarization and Source Code Summarization}

The problems close to our research problem are text summarization and source code summarization.
In particular,
text summarization aims to summarize the text documents, which can obtain a brief overview of a large text document.
While source code summarization can automatically generate the corresponding code comments by analyzing the semantic information of the target code, which can help developers understand the design purpose and functionality of the code.
Recently, 
the deep learning-based method is the popular research direction of these two research problems.
In this subsection, we mainly analyze the related work for these two problems.

For deep learning-based source code summarization, 
Iyer et al.~\cite{iyer2016summarizing} first studied the problem of source code summarization problem and proposed the method CODE-NN. This method mainly uses LSTMs in the encoder and decoder. At the same time, it used an attention mechanism, which can assign higher weights to related word elements in the sequence.
Hu et al.~\cite{hu2018deep} proposed the method DeepCom. Then they further extended the method DeepCom and proposed the method Hybrid-DeepCom~\cite{hu2020deep}, which combines the lexical information and grammatical information of the code, and splits the identifiers based on the camel case naming convention to alleviate the OOV problem. Finally, they used beam search to improve comment quality.
Yang et al.~\cite{yang2021comformer} proposed a novel method ComFormer based on Transformer and fusion method-based hybrid code presentation
LeClair et al.~\cite{leclair2020improved} considered the graph neural network, based on the graph2seq model~\cite{xu2018graph2seq}.
Ahmad et al.~\cite{ahmad2020transformer} considered the Vanilla-Transformer architecture, and used a relative position representation and copy mechanism.
Feng et al.~\cite{feng2020codebert} proposed the bi-modal pre-training model CodeBERT based on Transformer neural architecture for programming language and natural language.
Wang et al.~\cite{wang2020fret} proposed a BERT-based functional enhanced transformer model, they proposed a new enhancer to generate higher-quality code summarization.

For deep learning-based text summarization,
Rush et al.~\cite{rush2015neural} proposed a fully data-driven approach to abstracting sentence summaries. This approach used a local attention-based model to generate each word of the summary conditioned on the input sentence.
See et al.~\cite{see2017get} introduced a pointer network to address the problem that seq2seq models often do not accurately reproduce factual details. The approach can both generate words from a vocabulary through generators and copy content from a source through pointers. 
Liu and Lapata~\cite{liu2019text} used a pre-trained Bert~\cite{devlin2018bert} as a sentence encoder and a Transformer as a document encoder. The classifier of sentence representation is used for sentence selection. It used the knowledge of fine-tuned BERT to generate better text summaries.
Instead of only pre-training the encoder, the Bart model proposed by Lewis et al.~\cite{lewis2019bart} jointly pre-trained a seq2seq model that combines a bidirectional encoder and an autoregressive decoder.

\subsection{Novelty of Our Study}

Most relevant to our study is the approach {\CodeQue} proposed by Gao et al.~\cite{gao2020generating}. 
{\CodeQue}~\cite{gao2020generating} is an LSTM-based sequence-to-sequence deep learning model that helps improve the post quality by automatically generating post titles based on a given code snippet. However, {\CodeQue} only considers a single input modality and ignores the problem description of the question post, which can provide valuable information for title generation that code snippets sometimes cannot provide. Therefore, our approach {\tool} extracts useful information from both modalities (i.e., the problem description and the code snippet) simultaneously to generate high-quality titles for question posts.
In addition, we formalize the generation of question post titles for each programming language as separate but related tasks, which can allow related tasks to improve each other's performance by using shared and complementary information.

\section{Our Proposed Approach}
\label{sec:approach}

\begin{figure*}[htbp]
	\centering
    \vspace{-1mm}
	\includegraphics[width=0.9\textwidth]{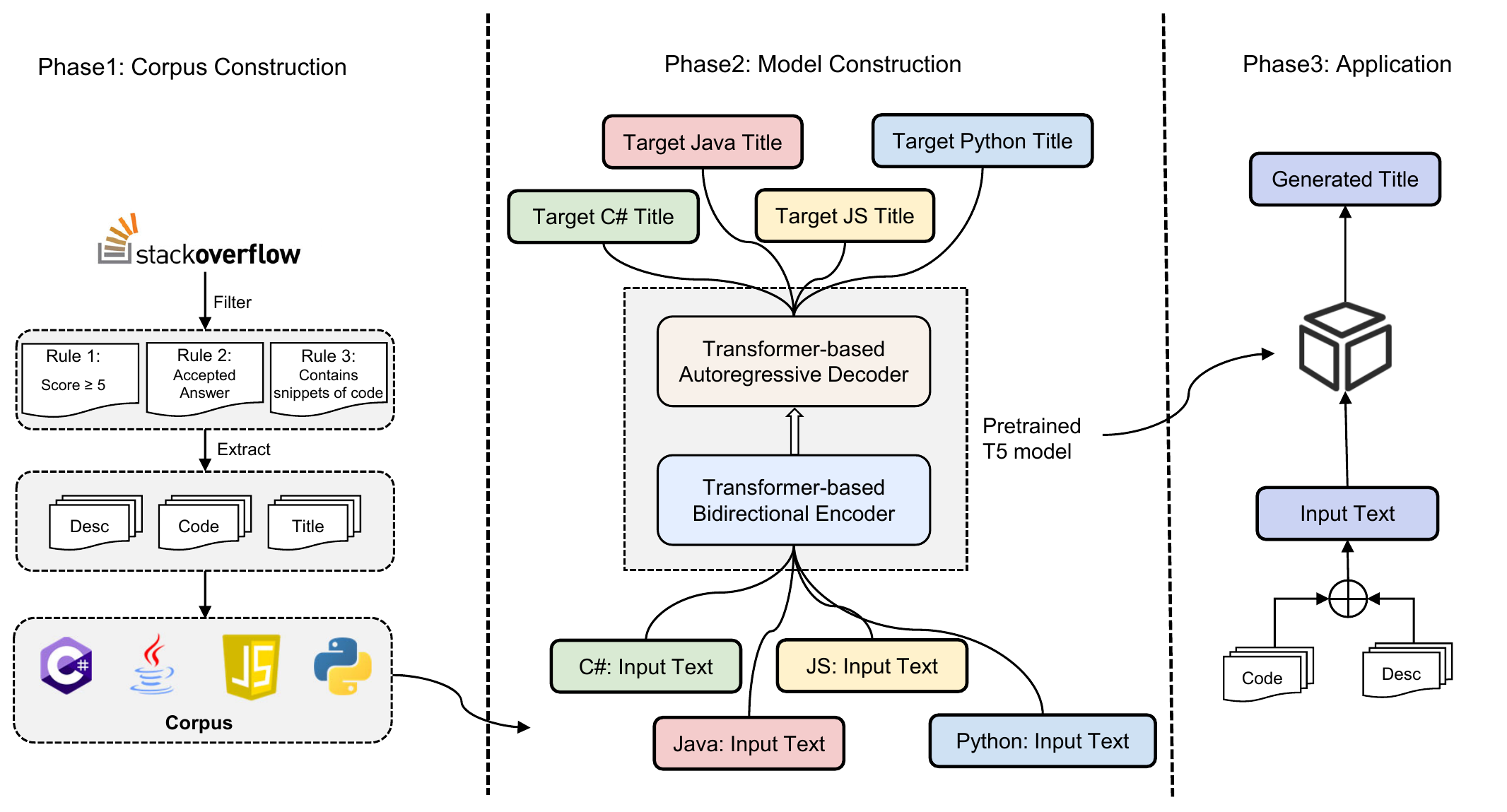}
	\caption{Framework of our proposed approach {\tool}}
    \vspace{-1mm}
	\label{fig:Framwork}
\end{figure*}

The overall framework of our approach is shown in \figurename~\ref{fig:Framwork}.
In this figure, we can find that {\tool} consists of three phases  (i.e., corpus construction, model construction, and model application). In particular, (1)
in the corpus construction phase, we design three heuristic rules to collect high-quality question posts from Stack Overflow. In our constructed corpus, we mainly focus on posts related to four programming languages and extract the code snippet, the problem description, and the title as a triplet from each post. More details of corpus construction can be found in Section~\ref{sec4:subject}.
(2) 
In the model construction,  we first model the multi-modal input by concatenating the code snippet and the problem description. Then we use the SentencePiece method~\cite{kudo2018sentencepiece} to split these two  modalities to alleviate the OOV problem.
Later we formalize question post title generation for each
programming language as separate but related tasks and resort to  multi-task
learning.
Finally, our model is fine-tuned based on a pre-trained Transformer  model T5~\cite{raffel2019exploring}.
(3)
In the model application phase, for a new question post, we input its code snippet and problem description to the constructed model. Then the trained model can automatically generate the corresponding post title through the beam search algorithm.
In the rest of this section, we show the details of the model construction phase.

\subsection{Multi-modal Input Modeling}

Since our study needs to deal with different programming languages simultaneously, we should alleviate the OOV problem.
In this study, we adopt the SentencePiece method~\cite{kudo2018sentencepiece} to handle our input $X$. The SentencePiece method considers the input sequence as a Unicode encoded sequence, thus this method does not depend on the language representation, which helps to handle different programming languages.

Different from previous studies, we want to extract information from both modalities (i.e., code snippet and problem descriptions) simultaneously. 
% There are two approaches to deal with these two modalities. The first approach is to feed the two modalities into two encoders separately. The second approach is to splice the code snippet and the problem description together and then feed them into one encoder. 
In this study, we   model the multi-modal input by concatenating the code snippet and the problem description together.
Specifically, we concatenate the code snippet sequence $X_{\text {code }}$ and the problem description sequence $X_{\text {desc }}$ via a special identifier ($<\operatorname{code}>$) to distinguish $X_{\text {code }}$ and $X_{\text {desc}}$.
As shown in \figurename~\ref{fig:Framwork}, since we considered generating titles for different programming language problem posts as separate but related tasks, we prefixed the input $X$ of each programming language with a task-specific prefix (e.g., the prefix ``JS:" denotes JavaScript) to make the model distinguish between the different tasks. The format of the input is shown as follows.

\begin{equation}
\mathrm{X}=\text { prefix }\oplus X_{\text {desc }} \oplus<\operatorname{code}>\oplus X_{\text {code }}
\end{equation}

\begin{comment}
由于我们的模型需要同时处理多种编程语言和自然语言的输入，因此OOV问题更加严重。为了缓解OOV问题，我们采用了sentencepiece来处理我们的输入。SentencePiece将序列看做unicode编码序列，从而子词算法不用依赖于语言的表示，有助于处理多种编程语言问题。

与以前的研究不同的是，我们要同时从代码片段和问题描述这两种模态中提取信息。处理这两种模态的方式有两种，第一种是将两种模态分别输入两个编码器中；另一种是将代码片段和问题描述进行拼接后输入到同一个编码器中。我们最终选择了第二种，详细细节参见RQ2. 具体来说，我们将代码序列Xcode和问题描述序列Xdesc直接进行拼接，使用特殊的标识符来区分Xcode和Xdesc，从而得到模型的输入X.
由于我们将为不同的编程语言问题帖子生成标题看作是不同但相关的任务，因此我们在每种编程语言的输入文本前添加特定于任务的前缀来让模型适应并区分不同的任务。

\end{comment}

\subsection{Transformer-based Bidirectional Encoder}

The encoder of our model aims to learn the representation of the problem posts $X=\left(x_{1}, x_{2}, \cdots, x_{m}\right)$. 
%Before feeding into the encoder, each token is embedded in a vector (i.e., $\boldsymbol{X}=\left(\boldsymbol{x}_{\boldsymbol{1}}, \cdots, \boldsymbol{x}_{\boldsymbol{m}}\right)$).
The encoder is composed of a stack of ``blocks", each of which consists of two subcomponents: a self-attention layer, followed by a feed-forward network.

Self-attention~\cite{ersulong} is calculated based on queries ($Q$), keys ($K$), and values ($V$). The dot product between the queries and keys is first calculated, then each is divided by $\sqrt{d_{k}}$ and the softmax function is applied to get the weight of the corresponding value.

\begin{equation}
\text { Attention }(Q, K, V)=\operatorname{softmax}\left(\frac{Q K^{T}}{\sqrt{d_{k}}}\right) V
\end{equation}

The feed-forward neural network (FFN) consists of two linear transformations, with a nonlinear transformation provided through the Relu activation.

\begin{equation}
\operatorname{FFN}(x)=\max \left(0, x W_{1}+b_{1}\right) W_{2}+b_{2}
\end{equation}

Layer normalization~\cite{ba2016layer} is applied to the inputs of each child component. After layer normalization, a residual skip connection~\cite{he2016deep} adds the inputs of each child component to its output. Dropout~\cite{srivastava2014dropout} is applied within the feed-forward network, on the skip connection, on the attention weights, and at the input and output of the entire stack.
%T5文章里的原话：First, an input sequence of tokens is mapped to a sequence of embeddings, which is then passed into the encoder. The encoder consists of a stack of “blocks”, each of which comprises two subcomponents: a self-attention layer followed by a small feed-forward network. Layer normalization (Ba et al., 2016) is applied to the input of each subcomponent. We use a simplified version of layer normalization where the activations are only rescaled and no additive bias is applied. After layer normalization, a residual skip connection (He et al., 2016) adds each subcomponent’s input to its output. Dropout (Srivastava et al., 2014) is applied within the feed-forward network, on the skip connection, on the attention weights, and at the input and output of the entire stack.

Note that the encoder uses all sub-tokens in the input sequence for learning so that the learned representation of each sub-token contains information about the entire input sequence. Since we encode the information of all the modalities in a sequence, the learned representation of each sub-token contains information about the other modalities.

\begin{comment}

编码器旨在学习问题帖主体的表示X=x1，x1，...，xn。在馈入编码器之前，每个token被嵌入到向量（即，X = （X=x1，x1，...，xn））中。然后将该序列传递给编码器。 
编码器由两个子组件组成：a self-attention layer followed by a small feed-forward network.
自我注意是根据查询、键和值计算的。首先计算查询与键之间的点积，然后给每一个键除以dk，并应用softmax函数获得相应值的权值。

Self-attention is calculated based on queries ($Q$), keys ($K$), and values ($V$).The dot product between the queries and keys is first calculated, then each is divided by $\sqrt{d_{k}}$ and the softmax function is applied to get the weight of the corresponding value.
\begin{equation}
\text { Attention }(Q, K, V)=\operatorname{softmax}\left(\frac{Q K^{T}}{\sqrt{d_{k}}}\right) V
\end{equation}

位置前馈网络由两个线性变换组成，通过Relu激活函数提供非线性变换。
\begin{equation}
\operatorname{FFN}(x)=\max \left(0, x W_{1}+b_{1}\right) W_{2}+b_{2}
\end{equation}

层归一化~\cite{ba2016layer}被应用于每个子组件的输入，在层标准化之后，a residual skip connection~\cite{he2016deep}将每个子组件的输入添加到其输出。Dropout~\cite{srivastava2014dropout}is applied within the feed-forward network, on the skip connection, on the attention weights, and at the input and output of the entire stack.

值得注意的是，编码器使用输入序列中的所有子令牌来学习，因此，每个子令牌的学习表示包含关于整个输入序列的信息。因为我们在一个序列中对所有信息模态进行编码，所以每个子令牌的学习表示都编码关于其他模态的信息。

\end{comment}

\subsection{Transformer-based Autoregressive Decoder}

The structure of the decoder is similar to the encoder. The difference is that it uses a standard attention mechanism to focus on the encoder output after each self-attentive layer. The self-attention mechanism in the decoder also uses a type of autoregressive or causal self-attention that allows the model to only focus on the past outputs.

%解码器的结构与编码器相似，不同之处在于它的每个自注意层之后都包括一个标准的注意机制来注意编码器的输出。解码器中的自我注意机制也使用一种自回归或因果的自我注意，它只允许模型关注过去的输出。

\subsection{Model Fine-tuning Process}

Our model is fine-tuned based on  a pre-trained Transformer language model $\mathrm{T}5$~\cite{raffel2019exploring}. 
During the fine-tuning process, we do not tune the parameters of the model's bias and LayerNorm.weight's weights, and then use the Adafactor optimizer~\cite{shazeer2018adafactor} to fine-tune the other parameters.

The input text $x$ is tokenized as $\left\{x_{1}, \cdots, x_{|x|}\right\}$ and encoded as the learned embedding $e^{x}=\left\{e_{1}^{x}, \cdots, e_{|x|}^{x}\right\}$.
The encoder takes $e^{x}$ as the input and outputs a joint representation of their context
$h=\left\{h_{1}^{x}, \ldots, h_{|x|}^{x}\right\}=\operatorname{Enc}\left(e^{x}\right)$.
The decoder then iterates on the previously generated token $y_{<j}$ via self-attention, the encoder outputs $h$ via cross-attention, and then predicts the probability of the next text token $P_{\theta}\left(y_{j} \mid y_{<j}, x\right)=\operatorname{Dec}\left(y_{<j}, h\right)$.

We train our model parameters $\theta$ by minimizing the negative log-likelihood of the target text tokens $y$  for a given input text $x$. For each task, the formula can be defined as follows.

\begin{equation}
\mathcal{L}_{\theta}^{\mathrm{Task}}=-\sum_{j=1}^{|y|} \log P_{\theta}\left(y_{j} \mid y_{<j}, x\right)
\end{equation}

Finally, the loss functions for these four different programming languages (i.e., Java, C\#, Python, and JavaScript) can be defined as follows.

\begin{equation}
\mathcal{L}_{\theta}= (\sum_{i=1}^{4} \mathcal{L}_{\theta}^{\mathrm{Task_{i}}})/4
\end{equation}

%数据集，收集过程，以什么规则筛选，然后进行预处理，预处理之后的数据的统计信息，长度，大小，训练及测试及怎么划分

\section{Experiment Setup}
\label{sec:setup}

In our empirical study, we want to answer the following three research problems (RQs):

\noindent\textbf{RQ1:} Can our proposed approach {\tool} outperform state-of-the-art baselines via automatic evaluation?

% \noindent\textbf{Motivation.} This RQ is designed to provide objective research to evaluate the overall performance of our methods {\tool}.

\noindent\textbf{RQ2:} What are the contribution of different input modalities for the performance of {\tool}?

% \noindent\textbf{Motivation.} This research problem aims to understand what specific improvements bi-modal information can bring to the work.

\noindent\textbf{RQ3:} Can our proposed approach {\tool} outperform the state-of-the-art post title generation approach  {\CodeQue} via human study?

% \noindent\textbf{Motivation.} As a way to help developers, we want to get actual feedback from users. Therefore, we provide a human study through RQ3.

In RQ1 and RQ3, we aim to show the competitiveness of {\tool} via automatic evaluation and human study.
In RQ2, we aim to analyze the effect of the multi-modal input modeling for {\tool}.

\subsection{Experimental Subject}
\label{sec4:subject}

\begin{figure}[htbp]
	\centering
    \vspace{-1mm}
	\includegraphics[width=0.5\textwidth]{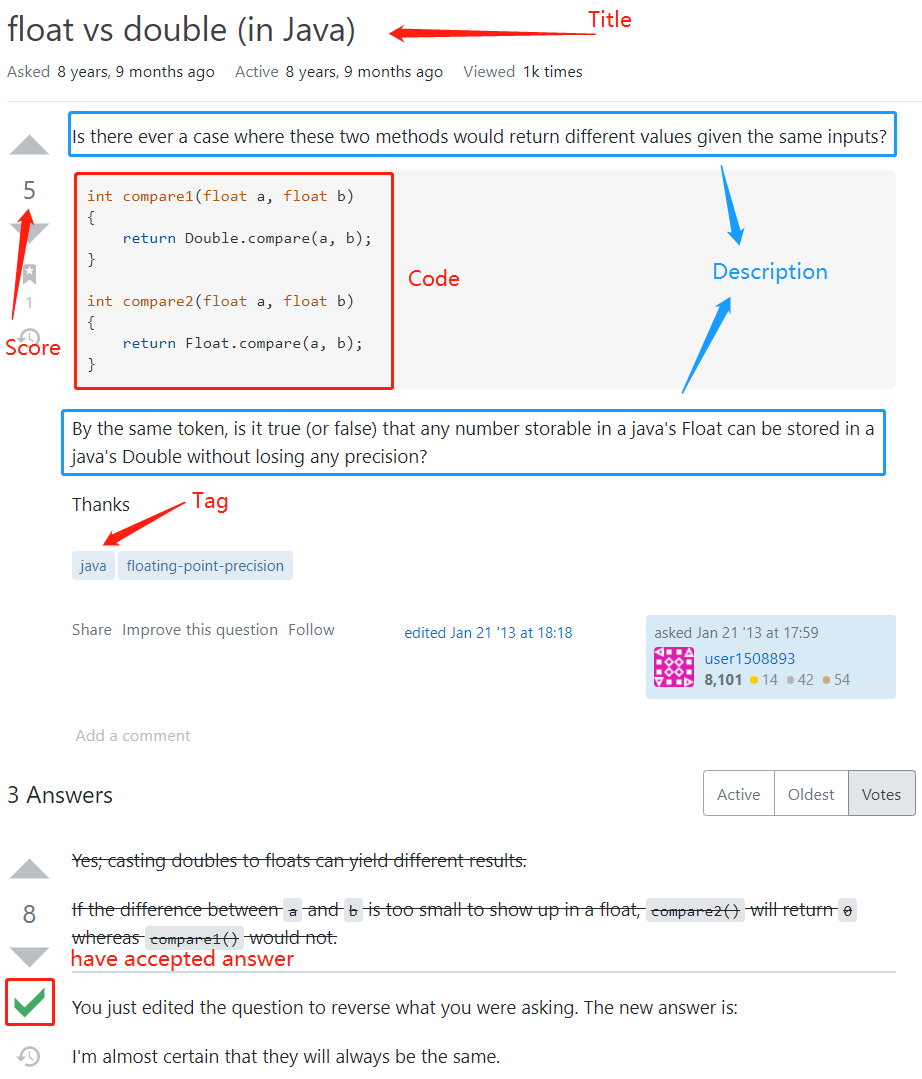}
	\caption{A question post for Java programming language from Stack Overflow}
    \vspace{-1mm}
	\label{fig:postExample}
\end{figure}

We select question posts from Stack Overflow as our experimental subject. 
\figurename~\ref{fig:postExample} shows a question post for the Java programming language.
This post contains a short post title, the problem description with the corresponding code snippet, one or more relevant answers (one of which was marked as accepted), and multiple tags.

In this study, we  mainly select question posts for four popular programming languages (i.e., Java, C\#, Python, and JavaScript).
Specifically, we first use Java, C\#, Python, and JavaScript tags to collect related question posts. 
To improve the quality of the gathered question posts, we propose three heuristic rules based on our manual analysis and suggestions from previous studies~\cite{bajaj2014mining}\cite{islam2019comprehensive}.

\begin{itemize}
    \item \textbf{Rule 1:} The score of the question posts is not smaller than 5.
    \item \textbf{Rule 2:} The question posts should have the accepted answers.
    \item \textbf{Rule 3:} The question posts should contain the code snippets.
\end{itemize}

After using these three rules, we find that the percentage of selected posts in Stack Overflow\footnote{\url{https://archive.org/download/stackexchange}, downloaded in October 2020} is only 6\%, which means we finally select 1,168,257 posts from 20,511,138 posts.
Then we extract the problem description, the code snippet, and the post title as the triplet $\langle$Description, Code, Title$\rangle$ and add this triplet to our corpus.
Finally, there are 68,959 posts for Java, 71,817 posts for C\#, 72,742 posts for Python, and 70,780 posts for JavaScript. 
Then we randomly selected 60,000 posts as the training set, 5,000 posts as the testing set, and the remaining posts as the validation set for each programming language.
Detailed statistical information of corpus split results can be found in Table~\ref{tab:CORPORA}. Moreover, we also show the length statistics of code snippet, problem description, and title in Table~\ref{tab:DatasetStatistics}.

% As shown in Table~\ref{tab:CORPORA}, there are 68,959 posts of Java, 71,817 posts of C\#, 72,742 posts of Python, and 70,780 posts of JavaScript in our constructed corpus.
% Then we randomly selected 60,000 posts for training and 5,000 posts for testing, with the rest for validation via stratified sampling. Moreover, we also show the length statistics of code snippet, problem description, and title in Table~\ref{tab:DatasetStatistics}.

\begin{table}[htbp]
  \centering
  \caption{Statistical information of corpus split results}
    \begin{tabular}{cccc}
    \toprule
    \textbf{Language} & \textbf{Training} & \textbf{Validation} & \textbf{Testing} \\
    \midrule
    Java  & 60,000 & 3,959 & 5,000 \\
    C\#   & 60,000 & 6,817 & 5,000 \\
    Python & 60,000 & 7,742 & 5,000 \\
    JavaScript & 60,000 & 5,780 & 5,000 \\
    \bottomrule
    \end{tabular}%
  \label{tab:CORPORA}%
\end{table}%

% Table~\ref{tab:DatasetStatistics} shows the length statistics of code snippet, problem description, and title.
% In this table, 
% we can find that the Length of Java code snippets is much longer than that of the other three programming languages. 
% On average, Java code snippets contain 173.98 tokens and JavaScript code snippets contain 142.93 tokens.
% The C\# and Python snippets are 128.54 and 124.98 in length, respectively.
% On the other hand, Java and C\# have somewhat longer problem descriptions than the other two programming languages, with Java and C\# averaging 114.09 and 120.15 respectively, and Python and JavaScript averaging 106.5 and 103.96 respectively.
% In addition, all programming languages have roughly the same title length, with an overall average title length of about 9 tokens.
% We found that more than 80\% of the code length in all programming languages was 256 tokens or less, more than 90\% of the problem descriptions were 256 tokens or less, and more than 90\% of the titles were 16 tokens or less.

\begin{table*}[htbp]
  \centering
  \caption{Length statistics of code snippet, problem description, and title}
    \begin{tabular}{ccccccccccccc}
    \toprule
    \multicolumn{1}{c|}{\multirow{2}{*}{\textbf{Language}}} &
      \multicolumn{4}{c|}{\textbf{Code Length}} &
      \multicolumn{4}{c|}{\textbf{Description Length}} &
      \multicolumn{4}{c}{\textbf{Title Length}} \\
    \multicolumn{1}{c|}{} &
      \textbf{Average} &
      \textbf{Mode} &
      \textbf{Median} &
      \multicolumn{1}{c|}{\textbf{\textless{}256}} &
      \textbf{Average} &
      \textbf{Mode} &
      \textbf{Median} &
      \multicolumn{1}{c|}{\textbf{\textless{}256}} &
      \textbf{Average} &
      \textbf{Mode} &
      \textbf{Median} &
      \textbf{\textless{}16} \\
    \midrule
   \multicolumn{1}{c|}{Java}  & 173.98 & 20    & 84   &  \multicolumn{1}{c|}{82.31\%} & 114.09 & 58    & 89    & \multicolumn{1}{c|}{93.45\%}  & 9.42  & 7     & 9     & 92.30\% \\
  \multicolumn{1}{c|}{C\#}   & 128.54 & 16    & 73    &\multicolumn{1}{c|}{88.40\%} & 120.15 & 62    & 93    &\multicolumn{1}{c|}{92.35\%}  & 9.69  & 8     & 9     & 90.90\% \\
   \multicolumn{1}{c|}{Python} & 124.98 & 24    & 70    &\multicolumn{1}{c|}{89.11\%}  & 106.55 & 48    & 85    &\multicolumn{1}{c|}{95.03\%}  & 9.41  & 8     & 9     & 92.86\% \\
   \multicolumn{1}{c|}{JavaScript} & 142.83 & 20    & 82    &\multicolumn{1}{c|}{86.68\%}  & 103.96 & 61    & 83    & \multicolumn{1}{c|}{95.34\%}  & 9.33  & 8     & 9     & 93.15\% \\
    \bottomrule
    \end{tabular}%
  \label{tab:DatasetStatistics}%
\end{table*}%

\subsection{Performance Measures}

In our study, we aim to show the competitiveness of {\tool} via both
automatic evaluation and human evaluation. In this section, we first introduce the performance measure used in the automatic evaluation. Then we introduce the details of our human study methodology.

\subsubsection{Automatic Evaluation}

Rouge~\cite{lin2004rouge} is a recall-based measure, which is used to calculate the lexical overlap between machine-generated summaries and reference summaries. 
This performance measure has been successfully used in previous source code summarization and text summarization studies~\cite{liu2019generating,raffel2019exploring,gros2020code}.
In particular, Rouge-1 and Rouge-2 are based on unigram and bigram respectively.
Rouge-L is based on the LCS (Longest Common Subsequence).
In our study, we use Rouge\footnote{\url{https://github.com/pltrdy/rouge}} to calculate the Rogue value.

\begin{figure*}[htbp]
	\centering
    \vspace{-1mm}
	\includegraphics[width=1\textwidth]{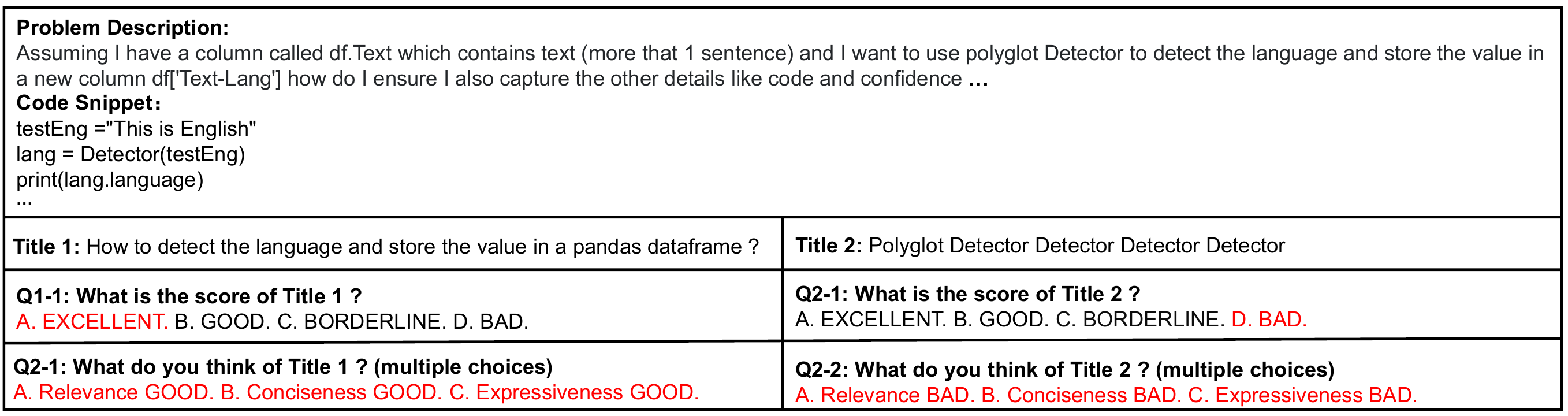}
	\caption{A survey example used in our human study}
    \vspace{-1mm}
	\label{fig:wenjuan}
\end{figure*}  
\subsubsection{Human Evaluation}

In our human study, we mainly follow the methodology of Chen et al.~\cite{chen2020stay}.
Specifically, we recruited six master students with more than five years on project development and familiar with the usage of Stack Overflow.
Before conducting the human study, we provide the guidelines for scoring the post title quality.
Since {\CodeQue}~\cite{gao2020generating} is the state-of-the-art baseline for post title generation, we mainly compare the quality of titles generated by {\tool} and {\CodeQue}.

% To collect real feedback from practitioners, we designed the human study. Since {\CodeQue}~cite{\gao2020generating} is the state-of-the-art baseline on this research problem, we mainly compare the quality of the titles generated by {\CodeQue} and our approach. Specifically, we followed the human study methodology of Chen et al.~\cite{chen2020stay} and recruited six volunteers who were graduate students majoring in computer science with more than five years of programming experience and experience using Stack Overflow.
% Before conducting the human study, we provided the volunteers with the title quality score guidelines.

% Overall, we would like to further analyze whether our approach can generate better issue titles for low quality issues in Stack Overflow through manual evaluation.
Then we randomly select 50 question posts for each programming language. Given a question post, we use  {\CodeQue} and {\tool} to generate titles respectively.
In summary, we have 200 titles generated by {\CodeQue}, which can be represented by $T^{i}=\left\{t_{1}^{i}, t_{2}^{i}, \cdots, t_{200}^{i}\right\}$), and 200 titles generated by {\tool}, which were represented by $T^{j}=\left\{t_{1}^{j}, t_{2}^{j}, \cdots, t_{200}^{j}\right\}$.

We randomly ordered the selected question posts, and the students did not know which title was generated by our approach.
We use the body of $k$-th question post and the associated $t_{k}^{i}$ and $t_{k}^{j}$ and the two generated titles are shown in random order.
We asked students to carefully review each question post.
First, we asked them to rate each of the assigned titles, the meaning and the corresponding score as shown as follows:

\begin{itemize}
    \item \textbf{Excellent (2).} This title can be used as the post title without modification.

    \item \textbf{Good (1).} This title can reflect the main idea of the question post, but has some problems (such as incompleteness, repetition, or grammatical problem). Therefore, this title can be used as the post title with minor modifications.

    \item \textbf{Borderline (0).} This title can reflect the idea of the question post, but it lacks necessary details or is confusing. Therefore, this title can be used as the post with major modifications.

    \item \textbf{Bad (-1).} This title  cannot reflect the idea of the question post. Therefore, this title needs to be rewritten.

\end{itemize}

By comparing the final score of $T^{i}$ and $T^{j}$, we can verify whether {\tool} can generate higher-quality titles than {\CodeQue}.

Second, we asked the hired students to give reasons to support their scores.
To simplify this process, we only surveyed the advantages of the generated titles which are scored as ``Excellent" or ``Good"  and the disadvantages of the generated titles which are scored as ``Borderline" or ``Bad".
The reasons are summarized from three perspectives.

\begin{itemize}
    \item \textbf{Relevance.} The generated title can capture the main idea of the question post with correct and sufficient detail.
    \item \textbf{Conciseness.} The generated title does not contain  unnecessary information and uses a short sentence to show the main idea of the post.
    \item \textbf{Expressiveness.} the generated title can be clearly described.
\end{itemize}

Here we use an  example post shown in \figurename~\ref{fig:wenjuan} from Stack Overflow\footnote{URL: \url{https://stackoverflow.com/questions/45958129}} to show the investigation process. With different answers in Q1, the options in Q2 are slightly different: (1) When the answer of Q1 is ``Excellent" or ``Good", the options of Q2 will ask the students why it is good, such as Q2-1. (2) When the answer of Q1 is ``Borderline" or ``Bad", the options of Q2 will ask students why it is bad, such as Q2-2.

\subsection{Implementation Details}

In our empirical study, we use Transformers\footnote{\url{https://github.com/huggingface/transformers}} to implement our proposed approach {\tool}.
The word embedding dimensions and hidden sizes are set to 768, and the number of attention heads and layers is set to 12.
All parameters were optimized with Adafactor~\cite{shazeer2018adafactor}, and the initial learning rate is set to 0.0005.
During training, the batch size is set to 30. The maximum length of the encoder and decoder is set to 512 and 30 respectively.
% the maximum length of the decoder to 30.
To alleviate the overfitting problem, we adopted a Dropout mechanism and set the loss rate to 0.1.
Finally, we also use the early stop method~\cite{prechelt1998early} to further alleviate the overfitting problem, and the weights with the highest performance on the validation set is taken as the final parameter value of the neural network.

We run all the experiments on a computer with an Inter(R) Xeon(R) Silver 4210 CPU and a GeForce RTX3090 GPU with 24 GB memory. The running OS platform is Windows OS. 

\subsection{Baselines}

We select six state-of-the-art baselines to show the effectiveness of our proposed approach.
We first select a recent post title generation approach {\CodeQue} as the first baselines.

\noindent\textbf{Code2Que.} {\CodeQue}~\cite{gao2020generating} is the state-of-the-art baseline for post title generation.  {\CodeQue} automatically generates titles for question posts through an LSTM-based deep learning approach.

Then we select three baselines from the text summarization domain.

\noindent\textbf{BM25.}  {\BM}~\cite{robertson2009probabilistic} is a bag-of-word retrieval function, which can be used to estimate the relevance of documents for a given search query. This is  the information retrieval-based baseline.

\noindent\textbf{Transformer.}  {\Transformer}~\cite{vaswani2017attention} mode improves the sequence processing ability through the attention mechanism. Recently, Transformer  has been successfully used in different NLP understanding and generation tasks.

\noindent\textbf{BART.}  {\BART}~\cite{lewis2019bart} pre-trained the Transformer model by combining bidirectional encoder and autoregressive decoder. The BART model performs well on text summarization tasks. In our study, 
we used the {\BART} model and then fine-tuned this model for our task.

Finally, we select two baselines from source code summarization. 

\noindent\textbf{NMT.}  Jiang et al.~\cite{jiang2017automatically} used Neural machine translation (NMT) technology to automatically ``translate" code changes into commit messages. We choose {\NMT} as a baseline since it can achieve promising performance in generating commit messages for the code changes. 

\noindent\textbf{Hybrid-DeepCom.}  {\DeepCom}~\cite{hu2020deep}  is a neural model used for code comment generation, which learns syntactic and semantic information about the code through two different encoders, respectively. It should be noticed that the structure of both {\NMT} and {\DeepCom} is the same except for the encoder part. We use the model {\DeepCom}  to generate post titles by learning code snippet and problem description separately through two different encoders. 
% We want to show that using one encoder for bi-modal coding is better than using two encoders separately for bi-modal coding by comparing the performance of {\DeepCom} with that of {\NMT}.

To make a fair comparison, we slightly adapted these baselines to deal with two modalities. In particular, 
for the baseline {\DeepCom}, we feed the two modalities into two encoders separately according to their original approach description. For the remaining baselines, we  concatenate the code snippet and the problem description
together and then feed them into one encoder.

\subsection{Browser Plug-in Support}

\begin{figure*}
	\centering
	\includegraphics[width=0.9\textwidth]{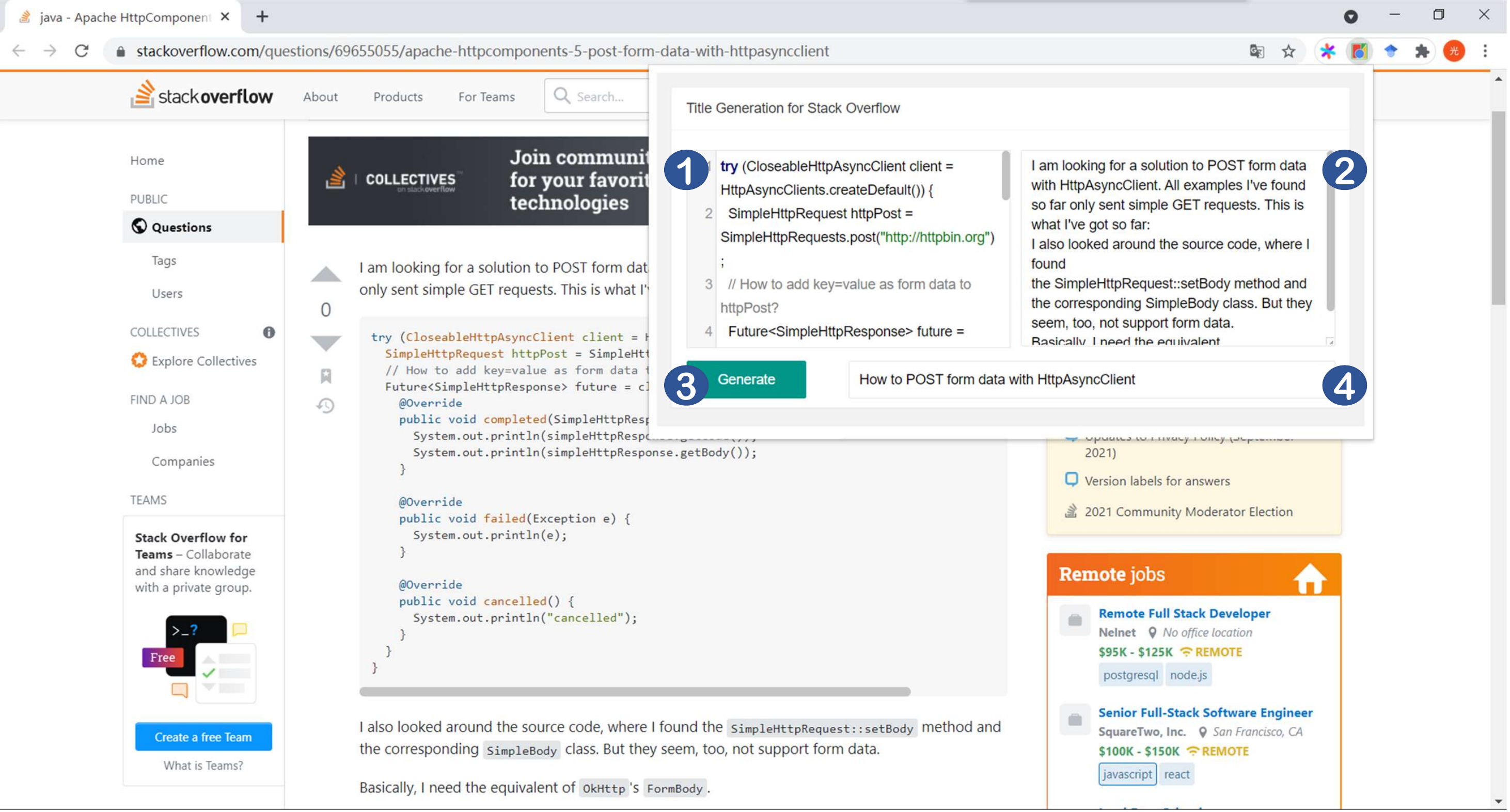}
    \vspace{-1mm}
	\caption{Screenshot of our developed browser plug-in. When using this plug-in, the developer can first copy the code snippet in \ding{172} and the detailed problem description in \ding{173}. Then the developer can click the generate button in \ding{174}. Finally, the generated title can be displayed in \ding{175}.}
    \vspace{-1mm}
	\label{fig:homepage}
\end{figure*}

To facilitate the use of our proposed approach, we developed a browser plug-in based on our constructed model. 
\figurename~\ref{fig:homepage} shows the screenshot of our developed browser plug-in. 
% \deleted{When using this plug-in, the developer can first copy the code snippet in \ding{172} and the detailed problem description in \ding{173}. Then the developer can click the generate button in \ding{174}. Finally, the generated title can be displayed in \ding{175}.}

\section{Result Analysis}
\label{sec:results}

\subsection{Result Analysis for RQ1}

\noindent\textbf{RQ1: Can our proposed approach {\tool} outperform state-of-the-art baselines via automatic evaluation?}

The comparison results between our proposed approach {\tool} and baselines can be found in Table~\ref{tab:RQ1}.
For each column, we emphasize the best value in bold. It can be found that our proposed approach {\tool} can significantly outperform six baselines in terms of  the performance measure Rouge for different programming languages.

In particular, for our considered baselines, we find deep learning-based baselines (i.e., {\NMT}, {\DeepCom}, {\CodeQue}, {\Transformer}, {\BART}, {\tool}) can achieve better performance than information retrieval-based baseline (i.e., {\BM}). The information retrieval-based baseline aims to retrieve historical similar posts based on similarity, and the performance of this kind of baseline  depends on whether  similar posts can be found. In contrast, deep learning-based approaches can learn
semantic information from two different modalities.

In terms of Rouge performance measures, our approach can significantly outperform the post title generation baseline {\CodeQue}~\cite{gao2020generating}.
For example, in terms of Rouge-L, compared to {\CodeQue}, {\tool} can improve the performance by 6.187\%, 6.005\%, 5.029\%, and 5.955\% for Java, Python, C\#, and JavaScript respectively.
The potential reason is that we consider the post title generation problem for different programming languages as different but related tasks, which can improve each other's performance by sharing information and complementing each other. 

Except for the Rouge performance measures, we also compare {\tool} with baselines in terms of METEOR~\cite{banerjee2005meteor} and BLEU~\cite{papineni2002bleu} performance measures.
Final results also show the competitive performance of {\tool}. Limited by the length of the paper, we show the detailed results on our project homepage.

\begin{table*}[htbp]
  \centering
  \caption{Comparison results between our proposed approach  and state-of-the-art baselines}
  \begin{center}
       \setlength{\tabcolsep}{1mm}{
 \resizebox{1\textwidth}{!}{
    \begin{tabular}{ccccccccccccc}
    \toprule
    \multicolumn{1}{c|}{\multirow{2}{*}{\textbf{Approach}}} &
      \multicolumn{3}{c|}{\textbf{Java}} &
      \multicolumn{3}{c|}{\textbf{C\#}} &
      \multicolumn{3}{c|}{\textbf{Python}} &
      \multicolumn{3}{c}{\textbf{JavaScript}}  \\
    \multicolumn{1}{c|}{} &
      \textbf{Rouge-1} &
      \textbf{Rouge-2} &
      \multicolumn{1}{c|}{\textbf{Rouge-L}}&
      \textbf{Rouge-1} &
      \textbf{Rouge-2} &
      \multicolumn{1}{c|}{\textbf{Rouge-L}}&
      \textbf{Rouge-1} &
      \textbf{Rouge-2} &
      \multicolumn{1}{c|}{\textbf{Rouge-L}}&
      \textbf{Rouge-1} &
      \textbf{Rouge-2} &
      \textbf{Rouge-L}\\
    \midrule
    \multicolumn{1}{c|}{BM25}  & 10.045  & 1.150  & \multicolumn{1}{c|}{9.294} & 10.183  & 1.760  & \multicolumn{1}{c|}{9.552}  & 11.443  & 1.475  & \multicolumn{1}{c|}{10.487}   & 9.782  & 1.182  & 9.205  \\
    \multicolumn{1}{c|}{NMT}   & 18.017  & 3.393  & \multicolumn{1}{c|}{17.166}  & 18.656  & 4.885  & \multicolumn{1}{c|}{17.850} & 19.482  & 4.090  & \multicolumn{1}{c|}{18.549}   & 17.341  & 3.279  & 16.738  \\
   
    \multicolumn{1}{c|}{Transformer} & 12.689  & 1.957  & \multicolumn{1}{c|}{12.332}  & 12.252  & 2.046  & \multicolumn{1}{c|}{12.115} & 14.125  & 2.236  & \multicolumn{1}{c|}{13.602}   & 13.153  & 2.082  & 12.751  \\
    \multicolumn{1}{c|}{BART} & 24.801  & 8.918  & \multicolumn{1}{c|}{23.343} & 23.395  & 9.545  & \multicolumn{1}{c|}{22.286} & 27.272  & 10.321  & \multicolumn{1}{c|}{25.672}    & 25.421  & 9.724  & 24.095  \\
     \multicolumn{1}{c|}{Hybrid-DeepCom}   & 10.620  & 1.511  & \multicolumn{1}{c|}{10.232}  & 12.346  & 2.684  & \multicolumn{1}{c|}{11.884} & 11.512  & 1.853  & \multicolumn{1}{c|}{11.156}   & 14.103  & 2.421  & 13.323  \\
    \multicolumn{1}{c|}{{\CodeQue}} & 22.021  & 6.787 & \multicolumn{1}{c|}{21.075}  & 23.476  & 8.605  & \multicolumn{1}{c|}{22.584} & 24.635  & 8.301  & \multicolumn{1}{c|}{23.282}   & 23.958  & 8.164  & 22.942  \\
    \multicolumn{1}{c|}{{\tool}} & \textbf{29.328 } & \textbf{10.968 } & \multicolumn{1}{c|}{\textbf{27.262}} & \textbf{29.550 } & \textbf{11.958 } & \multicolumn{1}{c|}{\textbf{27.613}} & \textbf{31.828 } & \textbf{11.985 } & \multicolumn{1}{c|}{\textbf{29.287}} & \textbf{31.150 } & \textbf{11.822 } & \textbf{28.897 } \\
    \bottomrule
 \end{tabular} } }
 \end{center}
  \label{tab:RQ1}
\end{table*}%

\begin{figure*}[htbp]
	\centering
    \vspace{-1mm}
	\includegraphics[width=1\textwidth]{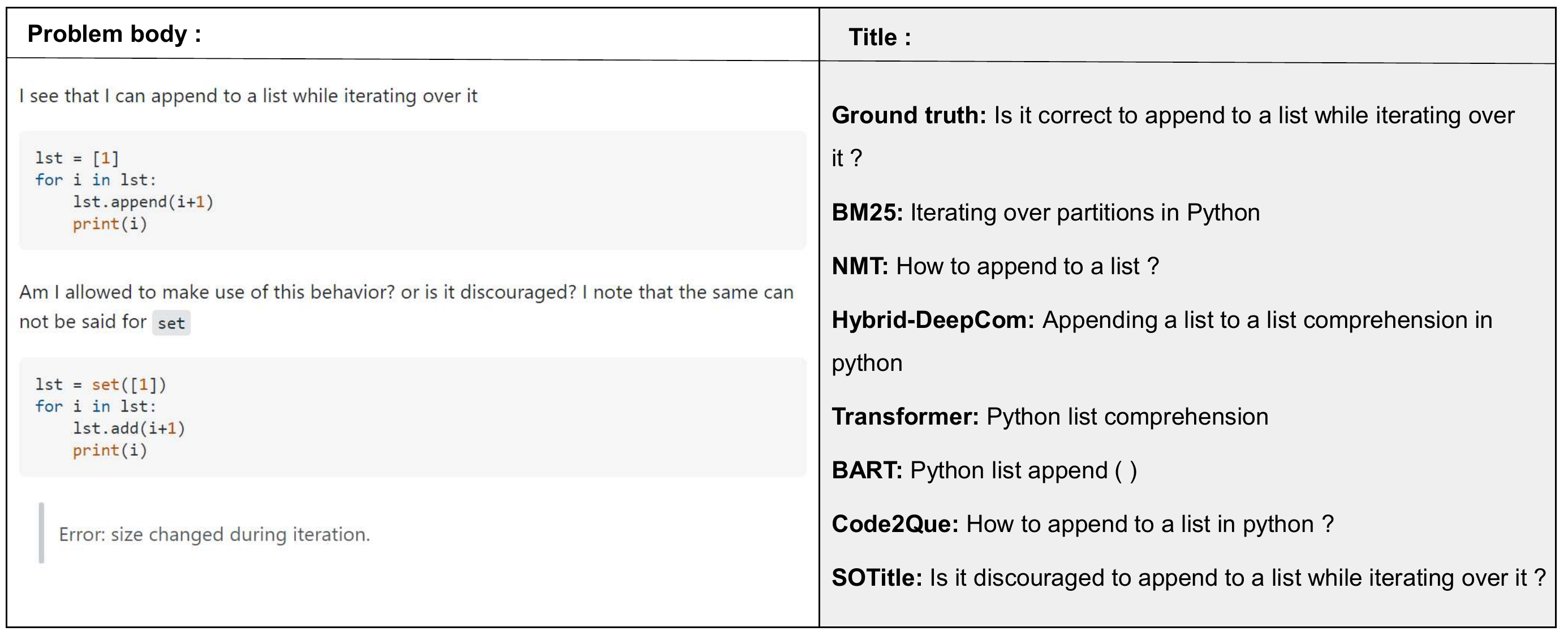}
	\caption{The titles generated by {\tool} and baselines for a question post related to the Java programming language}
    \vspace{-1mm}
	\label{fig:genCompare}
\end{figure*}

\figurename~\ref{fig:genCompare} shows the titles generated by {\tool} and baselines for a Java question post\footnote{URL: \url{https://stackoverflow.com/questions/52852143}}.
In this example, we can find
 that the title generated by {\BM} is not relevant to the question post. 
 The titles generated by  {\NMT}, {\DeepCom}.
 {\Transformer} and {\BART} cannot correctly express the core idea of the question post. {\CodeQue} fails to capture the details of the problem. However, the title generated by our approach  can accurately and fluently express the key information in this post.

\vspace{0.6cm}
\begin{tcolorbox}[width=1.0\linewidth, title={}]
\textbf{Summary for RQ1:} 
{\tool} can achieve better performance than six state-of-the-art baselines for different programming languages via automatic evaluation.
\end{tcolorbox}

\subsection{Result Analysis for RQ2}

\noindent\textbf{RQ2: What are the contribution of different input modalities for the performance of {\tool}?}

In this RQ, we aim to investigate the contribution of different input modalities to the performance of {\tool}. 
Moreover, we also investigate the contribution of different input modalities to the post title generation baseline {\CodeQue}, since {\CodeQue} only considers the code snippet in the original study.
Here we use different subscripts to distinguish different control approaches and the meaning of these subscripts is shown as follows.

\begin{itemize}
    \item \textbf{code.} The corresponding control approach only uses the code snippet as the input modal.
    \item \textbf{desc.} The corresponding control approach only uses the problem description as the input modal.
\end{itemize}

% In Section~\ref{sec:setup}, we use two input information in {\tool} (i.e., the code snippet and the problem description). At the same time, we use {\CodeQue} as the baseline model.
% Here we examine different combinations of these input modals.
% Specifically, we investigated the impact of these two sources of information: (1) code snippets and (2) problem descriptions.

Table~\ref{tab:RQ2} shows the performance of {\tool} and {\CodeQue} with different combinations of input modes.
Here {\CodeQue}$_{\text {code}}$ can denote the  performance of {{\CodeQue}}'s original study.
In terms of  Rouge-L performance measure, when we only use the code snippets as the input, the Rouge-L scores of {\tool} are 14.255, 14.255, 12.801, and 12.952 for the Java, C\#, Python, and JavaScript respectively; the Rouge-L scores of {\CodeQue} are 13.590, 13.602, 12.015, and 12.112 respectively.   
When we only use the problem description as the input, compared to only using the code snippets as the input. The Rouge-L scores of {\tool} can be improved by 61.9\%, 64.2\%, 97.1\%, and 92.4\% respectively; the Rouge-L scores of {\CodeQue} can be improved by 51.9\%, 61.2\%, 86.8\%, and 84.2\% respectively. This shows that for the post title generation task, more useful information can be extracted from the problem description. 
When we use both the problem description and the code snippet as the input, the Rouge-L scores of {\tool} can be improved by 91.2\%, 93.7\%, 128.8\%, and 123.1\% respectively. The Rouge-L scores of {\CodeQue} can be improved by 55.1\%, 66.0\%, 86.8\%, and 89.4\%, respectively.
This shows that the information from the code snippet and the problem description is complimentary and both two modalities should be considered for post title generation. 

\begin{table*}[htbp]
  \centering
  \caption{The performance of {\tool} and {\CodeQue} with different combinations of input modes}
  \begin{center}
       \setlength{\tabcolsep}{1mm}{
 \resizebox{1\textwidth}{!}{
    \begin{tabular}{ccccccccccccc}
    \toprule
    \multicolumn{1}{c|}{\multirow{2}{*}{\textbf{Approach}}} &
      \multicolumn{3}{c|}{\textbf{Java}} &
      \multicolumn{3}{c|}{\textbf{C\#}} &
      \multicolumn{3}{c|}{\textbf{Python}} &
      \multicolumn{3}{c}{\textbf{JavaScript}}  \\
    \multicolumn{1}{c|}{} &
      \textbf{Rouge-1} &
      \textbf{Rouge-2} &
      \multicolumn{1}{c|}{\textbf{Rouge-L}}&
      \textbf{Rouge-1} &
      \textbf{Rouge-2} &
      \multicolumn{1}{c|}{\textbf{Rouge-L}}&
      \textbf{Rouge-1} &
      \textbf{Rouge-2} &
      \multicolumn{1}{c|}{\textbf{Rouge-L}}&
      \textbf{Rouge-1} &
      \textbf{Rouge-2} &
      \textbf{Rouge-L}\\
    \midrule
    \multicolumn{1}{c|}{{\CodeQue}$_{\text {code }}$}  & 14.157  & 2.576  & \multicolumn{1}{c|}{13.590} & 14.895  & 3.388 & \multicolumn{1}{c|}{13.602}  & 12.705 & 2.184 & \multicolumn{1}{c|}{12.015} & 12.648 & 2.109 & 12.112\\
    \multicolumn{1}{c|}{{\tool}$_{\text {code }}$}  & 15.064  & 2.644  & \multicolumn{1}{c|}{14.255} & 15.445  & 3.525 & \multicolumn{1}{c|}{14.255}  & 13.534 & 2.236 & \multicolumn{1}{c|}{12.801} & 13.467 & 2.295 & 12.952\\
     \midrule
    \multicolumn{1}{c|}{{\CodeQue}$_{\text {desc }}$} & 21.213  & 6.042 & \multicolumn{1}{c|}{20.637} & 22.896  & 8.022  & \multicolumn{1}{c|}{21.957} & 24.012  & 8.001  & \multicolumn{1}{c|}{22.439} & 23.288  & 7.979  & 22.310  \\
    \multicolumn{1}{c|}{{\tool}$_{\text {desc }}$} & 24.586  & 9.101 & \multicolumn{1}{c|}{23.083} & 24.812  & 10.063  & \multicolumn{1}{c|}{23.439} & 26.966  & 10.408  & \multicolumn{1}{c|}{25.226} & 26.294  & 10.176  & 24.922  \\
    \midrule
    \multicolumn{1}{c|}{Code2Que} & 22.021  & 6.787 & \multicolumn{1}{c|}{21.075}  & 23.476  & 8.605  & \multicolumn{1}{c|}{22.584} & 24.635  & 8.301  & \multicolumn{1}{c|}{23.282}   & 23.958  & 8.164  & 22.942  \\
    \multicolumn{1}{c|}{\textbf{\tool}} & \textbf{29.328 } & \textbf{10.968 } & \multicolumn{1}{c|}{\textbf{27.262}} & \textbf{29.550 } & \textbf{11.958 } & \multicolumn{1}{c|}{\textbf{27.613}} & \textbf{31.828 } & \textbf{11.985 } & \multicolumn{1}{c|}{\textbf{29.287}} & \textbf{31.150 } & \textbf{11.822 } & \textbf{28.897 } \\
    \bottomrule
 \end{tabular} } }
 \end{center}
  \label{tab:RQ2}%
\end{table*}%

% Looking specifically at our findings from the examples, as shown in Table~\ref{fig:genCompare}, the overlap between code snippets and their corresponding titles is very low in all examples, which means that the model does not extract enough useful information from the code snippets. However, {\tool} can achieve  better performance than {\CodeQue} regardless of the modal case, which further illustrates the competiveness of our approach.

\vspace{0.6cm}
\begin{tcolorbox}[width=1.0\linewidth, title={}]
\textbf{Summary for RQ2:}
Problem description can provide more valuable information than the code snippet for post title generation. Moreover,
 considering two modalities together can improve the performance of {\tool} and {\CodeQue}.

\end{tcolorbox}

\subsection{Result Analysis for RQ3}

\noindent\textbf{RQ3: Can our proposed approach {\tool} outperform the state-of-the-art post title generation baseline {\CodeQue} via human study?}

\begin{figure}[htbp]
	\centering
    \vspace{-1mm}
	\includegraphics[width=0.5\textwidth]{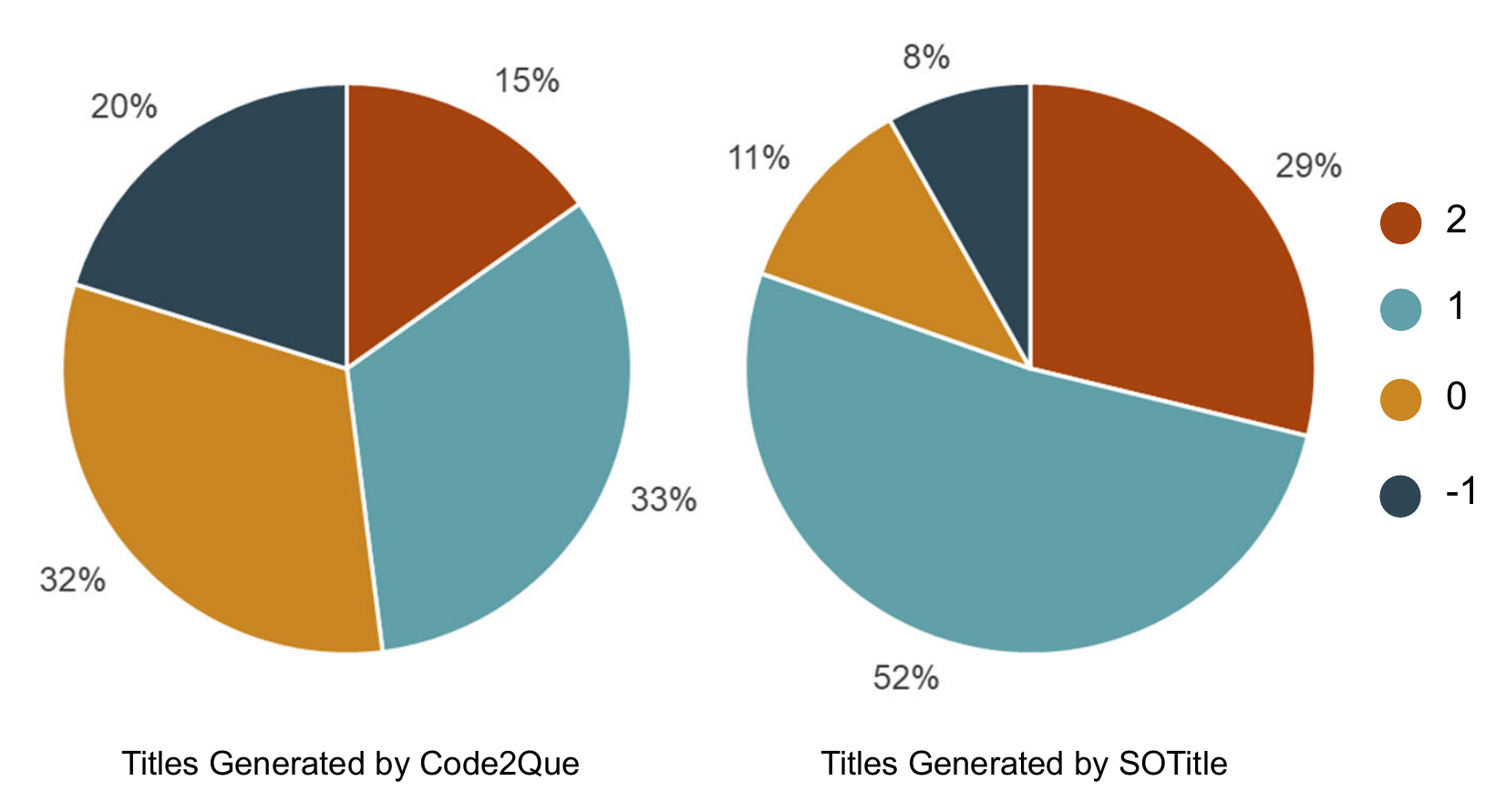}
	\caption{Rating distribution of human votes and each sector presents the proportion of votes with the corresponding rating score.}
    \vspace{-1mm}
	\label{fig:manualBing}
\end{figure}

\begin{figure}[htbp]
	\centering
    \vspace{-1mm}
	\includegraphics[width=0.5\textwidth]{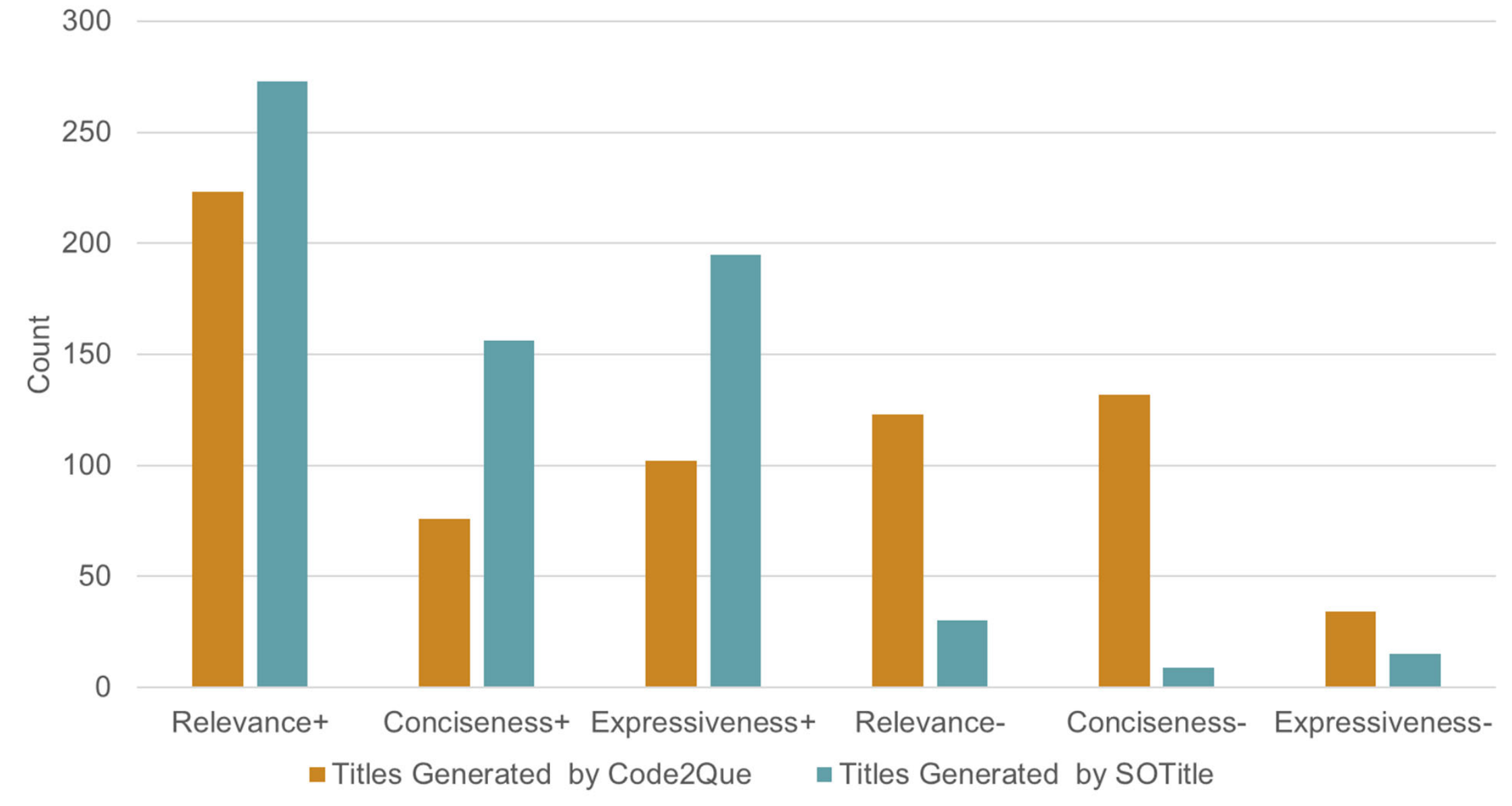}
	\caption{Advantages (+) and disadvantages (-) distribution of human comments. Each bar presents the count of corresponding perspective.}
    \vspace{-1mm}
	\label{fig:manualZhu}
\end{figure}

According to the methodology of the human study introduced in 
Section~\ref{sec:setup}, we randomly select 50 question posts for each programming language. For each post, two titles are generated by {\CodeQue} and {\tool} respectively. Each question post was evaluated separately by six master students during the survey. \figurename~\ref{fig:manualBing} and \figurename~\ref{fig:manualZhu} show the statistical results of 200 feedback comments, each containing four perspectives: Overall Rating, Relevance, Conciseness, and Expressiveness. 
% We calculate the percentage of user choices based on each evaluation metric.

As shown in \figurename~\ref{fig:manualBing}, the left and right subfigures show the votes for the titles generated by {\CodeQue} and the titles generated by {\tool}, respectively. We notice that 81\% of the human ratings for the  titles generated by  {\tool} are not less than 1, which means they are considered to have good quality and can be used as titles without major modifications. 
Then, we use Fleiss Kappa~\cite{fleiss1971measuring} to measure the agreement among these six students. The overall Kappa value based on the comparison results between {\tool} and {\CodeQue} is 0.766, which indicates substantial agreement among these students.
Moreover, the distribution of human comments (in \figurename~\ref{fig:manualZhu}) shows that the titles generated by {\tool} received more positive comments  after compared to the titles generated by {\CodeQue} in terms of three perspectives (i.e.,  Relevance, Conciseness, and Expressiveness). 
% This indicates that titles generated by {\tool} outperform the titles generated by {\CodeQue} from all the perspectives.
Therefore, our proposed approach can help to improve the post title quality.
% and that our approach generates clearer and more accurate issue headlines, which may help improve low-quality issues in Stack Overflow.
In our human study, we also find some negative votes for both approaches. After manual analysis, we find these approaches cannot generate useful titles when the code snippets are too complex or the problem descriptions lack enough useful information. We show the examples with negative votes and their corresponding reasons in our homepage.

\vspace{0.6cm}
\begin{tcolorbox}[width=1.0\linewidth, title={}]
\textbf{Summary for RQ3:} 
{\tool} can generate higher-quality titles than the baseline {\CodeQue} via human study.
\end{tcolorbox}

%\section{Discussions}
%\label{sec:discuss}

% \begin{figure*}[htbp]
% 	\centering
%     \vspace{-1mm}
% 	\includegraphics[width=1\textwidth]{figs/discuss.png}
% 	\caption{The same code has different titles}
%     \vspace{-1mm}
% 	\label{fig:discuss}
% \end{figure*}   

\section{Threats to validity}
\label{sec:threat}

In this section, we mainly analyze the potential threats to the validity of our empirical study.

\noindent\textbf{Internal Threats.} 
The first threat to internal validity concerns potential faults in the implementation of our proposed approach and baselines. To alleviate this threat, we use mature frameworks and use software testing to guarantee the code quality. 
% Moreover, we share our scripts and models with other researchers to facilitate and follow our study. 
The second threat to internal validity is the  method used to split the tokens.
In our study, we consider the SentencePiece method. This method is a language-independent subword tokenizer
and detokenizer, and its effectiveness has been verified in the neural machine translation problem~\cite{kudo2018sentencepiece}.

\noindent\textbf{External Threats.} 
The threat to external validity concerns the quality and generalization ability of our constructed corpus.
To alleviate this threat, we mainly consider question posts for four popular programming languages and consider three heuristic rules to identify and remove low-quality posts.

\noindent\textbf{Construct Threats.} 
The threat to construct validity comes from human studies, which may introduce bias. To ensure that all students can correctly understand our questionnaire, we provided a tutorial before our human study.

\section{Conclusion}
\label{sec:conclusion}

In this study, we propose the Transformer-based approach {\tool} for automatic post title generation by leveraging the code snippets and the problem description (i.e., the multi-modal input) in the question post.
% Then,
% we formalize question post title generation for each programming language as separate but related tasks and utilize multi-task learning to construct the model.
% Finally, we the SentencePiece approach~\cite{kudo2018sentencepiece} to concatenate the code snippet and the problem description to alleviate the OOV problem.
To verify the effectiveness of our proposed approach, we gathered 1,168,257 high-quality problem posts for four popular programming languages from Stack Overflow.
Experimental results show the competitiveness of our proposed approach after comparing the six state-of-the-art baselines in terms of Rouge performance measures.
Moreover, we also conduct a human study to verify the effectiveness of {\tool} after comparing the post title generation approach  {\CodeQue}.

In the future, we first aim to further improve the performance of {\tool} by considering more advanced deep learning methods (such as graph neural networks for code snippets). 
We second aim to combine {\tool} with other kinds of approaches (such as information retrieval-based approaches and template-based approaches). 
% In the future, we will further analyze the strengths and weaknesses of {\tool} through larger-scale experiments and delve into the key features of generating high-quality issue headers to improve our approach.

\section*{Acknowledgment}
The authors would like to thank the anonymous reviewers for their insightful comments and suggestions, which can substantially improve the quality of this work. 
Ke Liu and Guang Yang have contributed equally to this work and they are co-first authors.
This work is supported in part by the National Natural Science Foundation of
China (Grant no. 61872263), The Open Project of State Key Laboratory of Information Security (Institute of Information Engineering, Chinese Academy of Sciences) (Grant No. 2020-MS-07).

\bibliographystyle{IEEEtran}
\bibliography{mylib}
\end{document}